# Sampled Together: Assessing the Value of Simultaneous Co-located Measurements

C.E. Powell,[a,b] Christopher S. Ruf,[a] Scott Gleason,[c] Scot C. R. Rafkin [d]

[a] *Department of Climate and Space Sciences and Engineering,*
*University of Michigan, Ann Arbor, MI*

[b] *National Environmental Satellite Data and Information Service,*
*National Oceanic and Atmospheric Administration, Silver Spring, MD*

[c] *Daaxa LLC, Boulder, CO*

[d] *SouthWest Research Institute, Boulder, CO*

*Corresponding author*: C. E. Powell, cepowell@umich.edu

# ABSTRACT

This work applies a quantitative metric well-known to the data assimilation community to a new context in order to capture the relative representativeness of non-simultaneous or non-co-located observations and quantify how these observations decorrelate in both space and time. This methodology allows for the effective determination of thresholding decisions for representative matchup conditions, and is especially useful for informing future network designs and architectures.

Future weather and climate satellite missions must consider a range of architectural trades to meet observing requirements. Frequently, fundamental decisions such as the number of observatories, the instruments manifested, and orbit parameters are determined based upon assumptions about the characteristic temporal and spatial scales of variability of the target observation. With the introduced methodology, representativity errors due to separations in space and time can be quantified without prior knowledge of instrument performance, and errors driven by constellation design can be estimated without model ingest or analysis.

# 1. Introduction

It is often desirable to measure the Earth system from two or more different instruments at the same place and time. Simultaneous co-located measurements, otherwise known as matchups, frequently form the basis for calibration activities, science investigations, and operational retrievals. Satellite platforms offer a unique opportunity to capture co-located and simultaneous observations for extended periods of time. Many operational missions manifest multiple sensors onto the same satellite platform. Operational sea surface altimetry missions generally employ a radar altimeter as the primary mechanism to determine sea surface height, but due to uncertainties due to tropospheric delay, the radar is supplemented by a microwave radiometer to measure integrated atmospheric refractivity due to water vapor in the observing column to meet accuracy requirements (Donlon et al. 2021). Weather satellites utilize similar techniques to meet requirements. The NOAA-NASA Joint Polar Satellite System employs co-aligned microwave and infrared sounders to retrieve atmospheric profiles. To ensure operational consistency, the scanning mechanisms between the infrared and microwave instruments are synchronized such that they share the same field of regard across the scan (Kim et al. 2014).

The same logic also extends to formation flying, where instruments are not manifested on the same platform, but rather on multiple platforms in nearby, coordinated orbits. In the early 2000s, NASA populated its A-Train constellation satellites, named for its compact assemblage of several Earth science missions in the afternoon sun-synchronous polar orbit. When CloudSat and CALIPSO joined the A-Train in 2006, five separate satellites would fly in formation over the same ground track within roughly a 15 minute window (Stephens et al. 2002; Schoeberl 2002). The quick succession of satellites and near-simultaneous observations were critical to several science goals of the constellation.

The decision to co-manifest instruments on a single satellite platform usually involves various trade studies to evaluate the relative costs, risks, and performance benefits of the design. While sharing two or more instruments on the same satellite platform is often the most intuitive way to achieve simultaneity and co-location, it can increase the system complexity, as well as the volume, mass, and power budgets of the spacecraft. These budgets are known to drive overall mission cost and execution risk. At the other end of the spectrum, recent advances in miniaturized sensors, small satellite platforms, and low-cost launch services have enabled constellations of proliferated sensors. These new capabilities enable constellation designs previously considered untenable or uneconomical. There are inherent challenges and risks with proliferated constellations, including cross calibration, formation maneuvers, and operating complexity.

The transformations in the space industry, including the development of new business models for collecting observations from space, combined with growing demand for enhanced weather and climate services, are fostering new conditions for government agencies to consider as they embark on the next generation of Earth observing architectures. NOAA is currently formulating a new architecture for low Earth orbit, and is considering constellations that look very different than its legacy missions (Werner 2023). NASA recently commissioned a study from the National Academies of Science, Engineering, and Medicine to assess the utility of hosting a number of Earth science payloads on a single large commercial platform (National Academies of Sciences, Engineering, and Medicine 2023).

A number of tools have been developed to facilitate constellation design, usually with defined objective functions, such as minimizing revisit time, maximizing coverage, or balancing cost and utility (Marcuccio et al. 2019; Williams, Crossley, and Lang 2001; St. Germain, Gallagher, and Maier 2018; Nag et al. 2015). These considerations feature prominently in the development of Earth observing missions and frequently trade off with one another. However, often the most important questions for selecting an optimal architecture – the definition of threshold and objective revisit requirements for a given observation – are decided somewhat arbitrarily.

One of the primary metrics used to measure the utility of future observing systems, particularly for weather satellites, is a technique known as Observing System Simulation Experiments (OSSEs) (Arnold and Dey 1986). OSSEs essentially use a high-resolution model of the atmosphere and Earth system as ground truth, which are then "measured" by realistic simulated observations that are assimilated into a known forecasting model (Hoffman and Atlas 2016). These experiments can provide significant insight into the forecast impact of certain types of future observations but are quite labor intensive and computationally expensive to run and are usually individually tuned to the observing system being developed (to illustrate, cf. Li et al. 2019; 2018; Christophersen et al. 2021). But because of the number of variables inherent in these models, it is difficult to make precise decisions about what utility is gained by changing the temporal or spatial coverage of observations.

This work examines precisely how much representativity error is incurred when observations are separated in space and time, without any required *a priori* knowledge about the observing system at hand. As a result, it is a simple tool to empower observation planners with objective functions to make architecture trades.

## 2. Methodology

*A brief historical aside.*

The challenge of assimilating many different observations, which are often irregularly sampled in space and time, has been well documented since the advent of numerical weather prediction over a century ago (Richardson 1922). Gandin introduced a number of innovations in objective analysis and optimum interpolation of fields with statistical arguments for how parameters of interest might decorrelate in space and time (Gandin 1965), which was further generalized for data assimilation purposes by Rutherford (Rutherford 1972). Bretherton et. al. employed a substantially similar analysis for the design of an in-situ network and field campaign of ocean observations (Bretherton, Davis, and Fandry 1976).

For the purposes of this analysis, we will adopt Bretherton's formulation. The basic premise is that given simple assumptions about the statistical behavior of any observable parameter, such as wind speed or sea surface temperature, one can measure the rate that this parameter decorrelates over space and time. The error from interpolation is then simply:

$$\epsilon_x(\tau_{t,s}) = \sigma_x \sqrt{1 - R_x(\tau_{t,s})} \qquad (1)$$

where representativity error for parameter $x$ is $\epsilon_x$ and is a function of either time lag $\tau_t$ or spatial lag $\tau_s$, $\sigma_x$ is the standard deviation of samples of $x$, and $R_x$ is the autocorrelation of parameter $x$ at lag $\tau_{t,s}$. This formulation makes general assumptions about the stationarity and isotropy of variance for the parameter of interest which are not necessarily true for the Earth system. There are many cases in which the statistical distribution and rate of decorrelation of weather and climate parameters are non-stationary, such as from seasonality or phase of teleconnections. Generally, the error in (1) can be calculated in any case where a representative decorrelation roll-off can be estimated. The decorrelation behavior of Earth parameters is frequently studied and easily accessible for many parameters (Kuragano and Kamachi 2000; Chu, Guihua, and Chen 2002; McLean 2010; Eden 2007; Delcroix et al. 2005; Romanou, Rossow, and Chou 2006; Gille and Kelly 1996).

This analysis treats temporal and spatial decorrelation separately, even though they are coupled in the Earth system. The rationale is twofold. First, this maintains the simplicity and generality of the representativity errors. Second, this assumption makes this metric much more useful for the task of designing satellite architectures. Architecture trades frequently feature orbit and constellation decisions that optimize various sampling characteristics in both space and time, which are also coupled by orbital dynamics. Instead of trying to determine the dynamics of multiple coupled systems, this relaxation allows planners to set simple threshold and objective requirements in space and time for their trade studies.

*Temporal Decorrelation*

For this exercise, we will assume that we are designing an ocean observing system, with particular interest in surface parameters such as winds, sea surface temperature, air temperature, humidity, and surface air pressure. These parameters are readily available from moored marine buoys from NOAA's National Data Buoy Center (National Data Buoy Center 1971). Care should be taken to ensure that the source of "ground-truth" for target observables provide representative sampling at timescales appropriate for the phenomena of interest and are otherwise well calibrated to ensure representative variability and dynamic range. For our analysis, we examine one year of observations from Station 51004, located at 17°32'17" N 152°13'48" W (approximately 205 nm southeast of Hilo, HI), which is a 3-meter foam buoy with an updated SCOOP payload that reports meteorological parameters at 10-minute intervals (Kohler, LeBlanc, and Elliott 2015).

It is important to note that the choice of environmental parameter and source data may have a significant impact on the representativeness of decorrelation scales. Parameters such as precipitation are likely to have much shorter characteristic decorrelation scale sizes than sea surface temperature. It is assumed that the temporal decorrelation scales from Station 51004 data are sufficiently representative to use as a benchmark for determining satellite constellation objectives in this hypothetical. Station 51004 was chosen due to its long duration of continuous data collection without equipment changes as well as its near-equatorial location which highlights highly variable surface weather patterns. The goal of this exercise is to demonstrate how we apply a known benchmark to determine satellite constellations architectures, but in practice, care should be taken to ensure that these benchmarks are also representative of the target phenomenology.

The autocorrelation for observations was calculated assuming wide-sense stationarity for lags of 10 minutes. The decorrelation behavior and the resulting representativity error is shown in *Fig. 1*.

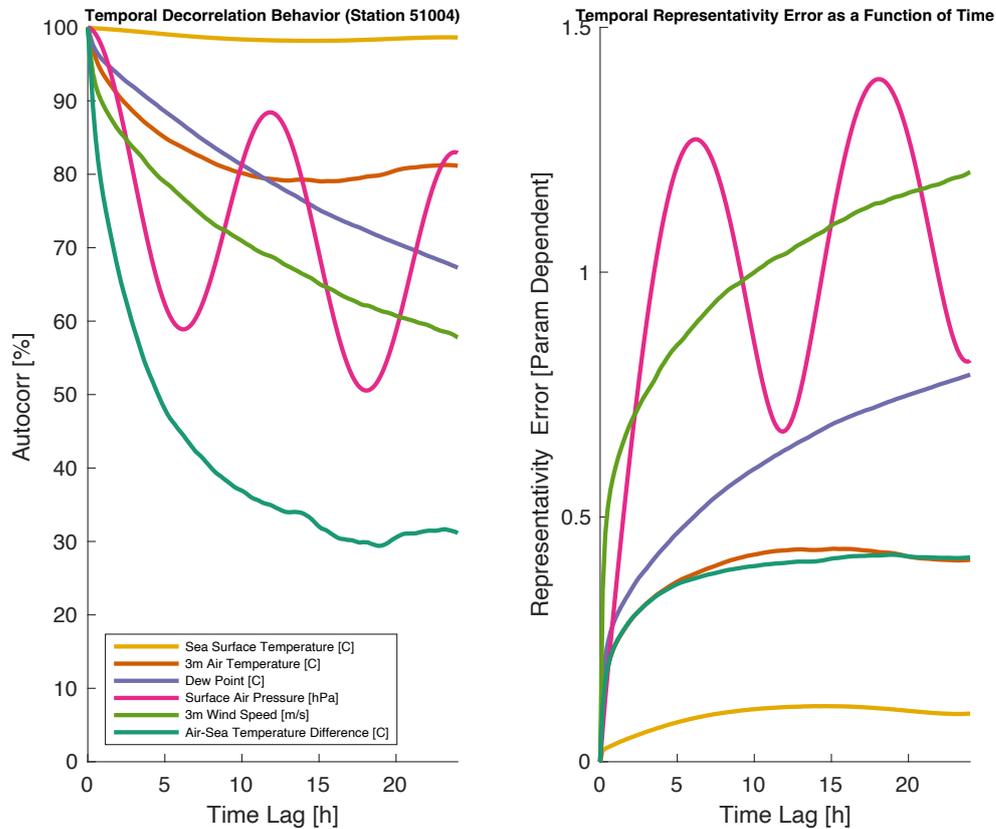

*Fig 1. (Left panel) The temporal decorrelation behavior for a year of observations is demonstrated for various meteorological parameters from NDBC buoy station 51004. (Right panel) The representativity error as calculated from (1) is shown. Note that the magnitude is dependent on the unit of measure.*

Figure 1 is illustrative of several factors that observation planners should consider. The left panel shows the decorrelation behavior of several observed parameters as well as one derived parameter, the air-sea temperature difference, which is obtained by simple arithmetic subtraction of the near surface air temperature from the sea surface temperature observations. Some environmental parameters, such as sea surface temperature, are slowly varying on the timescale of hours to days. Other parameters, such as surface air pressure, exhibit strong diurnal behavior. The air-sea temperature difference decorrelates at a much faster rate than either air temperature or sea surface temperature alone, suggesting that these parameters are decoupled on timescales of less than a day. The right panel shows how these decorrelation behavior factors into absolute representativity error. For example, the representativity error of air temperature and of the air-sea temperature difference are nearly equivalent, despite the fact that the decorrelation behavior is vastly different. This is because the air-sea temperature difference has a much smaller dynamic range than the surface air temperature. At time lags of 5 hours, both exhibit approximately 0.4º C of representativity error, but that value is much more significant for the air-sea temperature difference, which has a mean of -0.76º C and a standard deviation of 0.5º C.

It is important to emphasize that this formulation addresses only the component of matchup error due to representativity. The overall matchup error between two parameters should also include measurement and retrieval errors associated with the individual parameters. The representativity error can therefore be considered the floor for matchup accuracy of a given observing system. Engineers can design instruments that minimize measurement errors, but there is no amount of technology that can overcome fundamental errors due to the non-representativeness of the observation.

*Spatial Decorrelation*

A similar analysis can be applied in spatial dimensions. For a spatial dataset, we selected NASA's Modern-Era Retrospective analysis for Research and Applications version 2 (MERRA-2) hourly, non-averaged reanalysis (M2I1NXASM) (Gelaro et al. 2017; Global Modeling And Assimilation Office 2015). As before, we assume for this exercise that reanalysis data is sufficient to capture the spatial variability of our target parameters, but this assumption should be reconsidered for phenomena that vary at scale sizes smaller than the resolvability of the model.

Spatial decorrelation of ocean surface observations is known to be anisotropic and location dependent. This study adopts a simple solution frequently employed by oceanographers, which is to individually evaluate the meridional and zonal components of the spatial decorrelation scales of an ocean observation (White 1995; Reynolds and Smith 1994). For consistency, we center the meridional and zonal transects at the location of Station 51004 as illustrated in *Fig. 2*. While these statistics are not stationary across the ocean, we assume for the purposes of this hypothetical that the statistics from Station 51004 are representative.

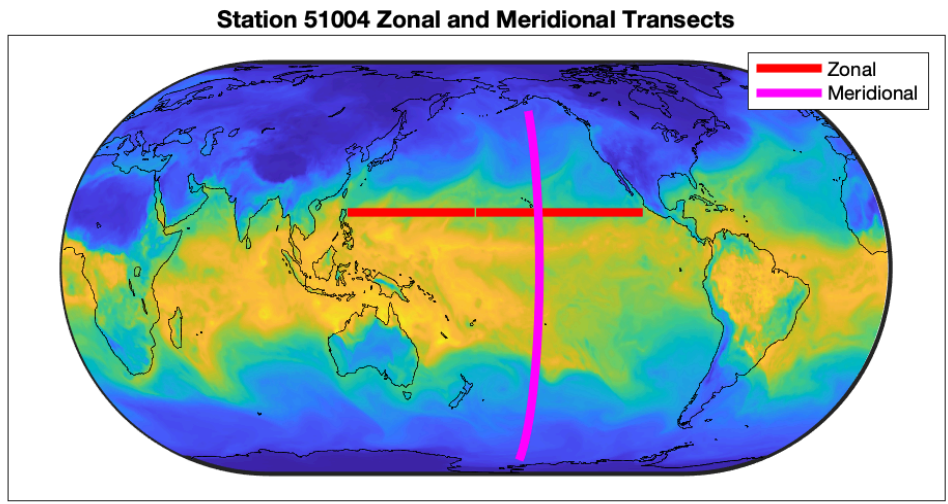

*Fig 2. The zonal and meridional transects for the Pacific basin centered at Station 51004 are highlighted. The zonal transect consists of 210 grid cells at 0.625 deg spacing. The meridional transect consists of 254 grid cells at 0.5 deg spacing. The background field is of an initialization of MERRA-2's global surface relative humidity output to emphasize that the spatial patterns of these parameters are quite different meridionally than they are zonally.*

Figure 2 identifies which model grid cells are used to compute the zonal and meridional autocorrelation statistics. Because there are only 210 data points in the zonal transect and 254 data points in the meridional transect, the autocorrelation behavior derived from a single model run is somewhat noisy. To create a more representative decorrelation behavior, the autocorrelation is averaged across multiple model realizations over three months of data (at each hourly output for 92 days, or for 2,208 realizations).

Compared to the NDBC buoy data, the MERRA-2 hourly reanalysis produces slightly different observation parameters. For instance, the air temperature is the 10 meter temperature, whereas the buoy data is observed at 3 meters. Additionally we use the MERRA-2 skin temperature, which over most of the oceans is very similar to the sea surface temperature. The MERRA-2 model also computes humidity in terms of relative humidity rather than dew point.

The average decorrelation behavior for the meridional and zonal transects centered at Station 51004, as well as the implied representativity error are shown in *Fig. 3*.

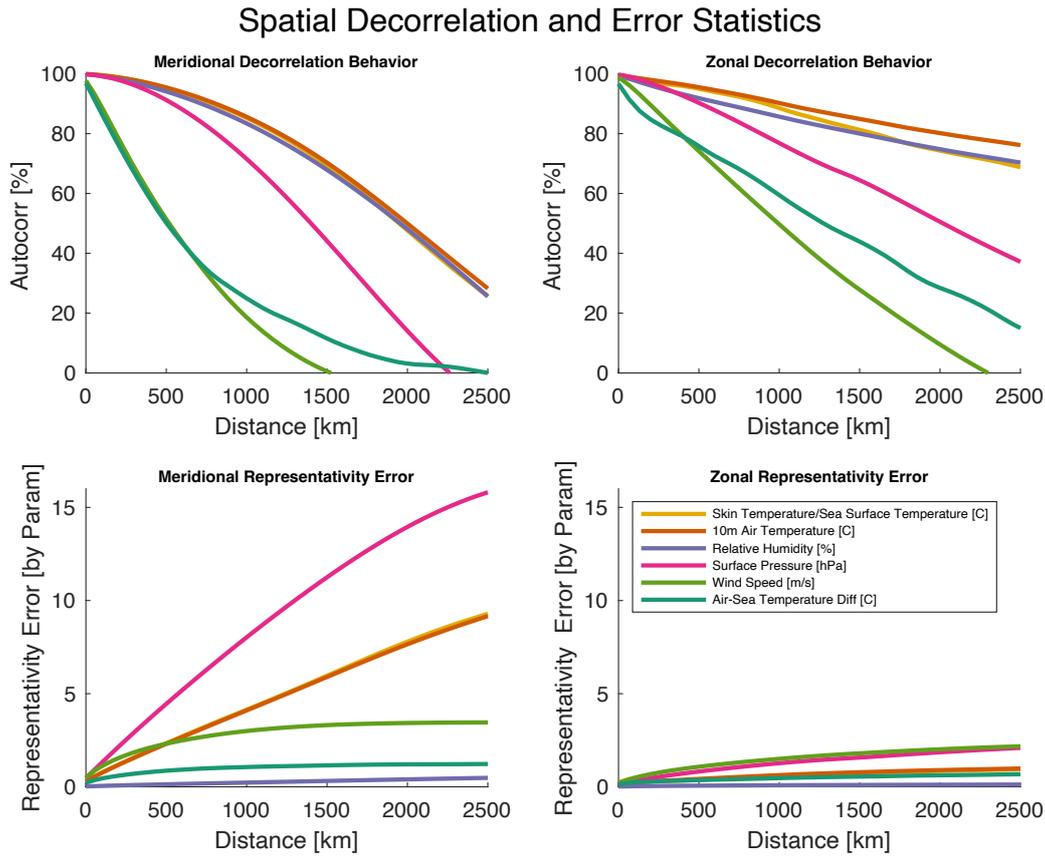

*Fig. 3. The spatial decorrelation behavior and error statistics are partitioned into meridional and zonal components. (Top left panel) The meridional decorrelation behavior is shown for the specified environmental parameters. (Top right panel) The zonal decorrelation behavior is shown for the specified environmental parameters. (Bottom left panel) The meridional representativity error is shown, note that the y-axis units are parameter-dependent. (Bottom right panel) The zonal representativity error is shown, note that the y-axis units are parameter-dependent.*

Figure 3 reveals much about the spatial decorrelation behavior of the selected parameters. First, the behavior is significantly different between zonal and meridional transects. Second, for the selected parameters, the spatial decorrelation roll-off is relatively slow, indicating characteristic spatial decorrelation scale sizes in the hundreds or even thousands of kilometers. Third, even when the decorrelation is comparable in zonal and meridional directions, the variability of the parameter may have significantly different magnitudes, changing the corresponding representativity error. For satellite missions, designers would want to base requirements decisions on the 'worst-case' scenario, which in this case, indicates that the meridional behavior and error statistics would generally drive architecture decisions.

## 3.    Discussion

It is worth repeating that the representativity error indicated in this analysis is not inclusive of the dominant error sources of most observing systems, such as ambiguity in the retrieval or noise from the sensor. In cases where the additional representativity error incurred by having non-simultaneous and non- co-located measurements is small relative to the dominant error sources, it becomes reasonable to consider more flexible architectural approaches. One of the advantages of using multiple smaller satellite platforms for Earth observations is that on-orbit resources are limited, and often the risk posture prohibits manifesting multiple instruments on the same observatory. This advantage can be traded off against the detrimental effects of representativity error. Further, the miniaturized sensors that enable this architecture often descope the extensive and expensive onboard calibration systems used by flagship national systems, thus potentially increasing their measurement errors.

As an example, consider the design of a constellation to observe ocean surface windspeed from two separate observatories with the same orbit ground track. Given a requirement of representativity error no greater than 0.5 m/s, the above results from Figure 1 indicate that the satellites be staggered no more than 25 minutes apart in the orbit plane. Note that this corresponds to an average representativity error under typical conditions. However, if the target observable is ocean winds for tropical cyclone monitoring, during which surface windspeed exhibits a much larger dynamic range and steeper decorrelation roll-off, a 25 minute separation would likely be too large.

This problem can be extended to thresholds established for matching up observations for opportunistic or vicarious calibrations. Techniques such as simultaneous nadir overpasses (SNOs) (Zou et al. 2006) often establish thresholds in space and time that approximate simultaneity. This technique can establish objective representativity guidelines which can optimize the quantity of matchup data, or enable constellations that feature frequent SNOs for operational calibration.

*Acknowledgments.*

This manuscript is the sole work of the listed authors and does not necessarily represent the views of the National Oceanic and Atmospheric Administration or the United States Department of Commerce.

This work was partially supported by NASA Science Mission Directorate contract NNL13AQ00C with the University of Michigan.

The authors declare no conflict of interest.